# The Third Evolution Equation for Optimal Control Computation

Sheng ZHANG, Fei LIAO, and Kai-Feng HE

(2018.01)

*Abstract:* **The Variation Evolving Method (VEM) that originates from the continuous-time dynamics stability theory seeks the optimal solutions with variation evolution principle. After establishing the first and the second evolution equations within its frame, the third evolution equation is developed. This equation only solves the control variables along the variation time to get the optimal solution, and its definite conditions may be arbitrary since the equation can eliminate possible infeasibilities. With this equation, the dimension of the resulting Initial-value Problem (IVP), transformed via the semi-discrete method, is greatly reduced. Therefore it might relieve the computation burden in seeking solutions. Illustrative examples are solved and it is shown that the proposed equation may produce more precise numerical solutions than the second evolution equation, and its computation time may be shorter for the dense discretization.**

## I. INTRODUCTION

Optimal control theory aims to determine the inputs to a dynamic system that optimize a specified performance index while satisfying constraints on the motion of the system. It is closely related to the engineering and has been widely studied [1]. Because of the complexity, Optimal Control Problems (OCPs) are usually solved with numerical methods. Various numerical methods are developed and generally they are divided into two classes, namely, the direct methods and the indirect methods [2]. The direct methods discretize the control or/and state variables to obtain the Nonlinear Programming (NLP) problem, for example, the widely-used direct shooting method [3] and the classic collocation method [4]. These methods are easy to apply, whereas the results obtained are usually suboptimal [5], and the optimal may be infinitely approached. The indirect methods transform the OCP to a Boundary-value Problem (BVP) through the optimality conditions. Typical methods of this type include the well-known indirect shooting method [2] and the novel symplectic method [6]. Although be more precise, the indirect methods often suffer from the significant numerical difficulty due to the ill-conditioning of the Hamiltonian dynamics, that is, the stability of the costates dynamics is adverse to that of the states dynamics [7]. The recent development, representatively the Pseudo-spectral (PS) method [8], blends the two types of methods, as it unifies the NLP and the BVP in a dualization view [9]. Such methods inherit the advantages of both types and blur their difference.

Theories in the dynamics and control field often enlighten strategies for the optimal control computation, for example, the non-linear variable transformation to reduce the variables [10]. Recently, a Variation Evolving Method (VEM), a dynamic method inspired by the states evolution within the stable continuous-time dynamic system, is proposed for the optimal control computation

The authors are with the Computational Aerodynamics Institution, China Aerodynamics Research and Development Center, Mianyang, 621000, China. (e-mail: zszhangshengzs1@outlook.com).

[11]-[18]. To the newest knowledge of the authors, it is found that actually early in the 1980s, Snyman has already proposed the dynamic method to solve the unconstrained Parameter Optimization Problems (POPs) [19], and applied it to the computation of OCPs [20], which are transformed to the unconstrained POPs with the penalty function and the control parameterization techniques. Differently, built upon the generalized Lyapunov principle, the VEM considers the infinite-dimensional dynamics. It also synthesizes the direct and indirect methods, but from a new standpoint. The Evolution Partial Differential Equation (EPDE), which describes the evolution of variables towards the optimal solution, is derived from the viewpoint of variation motion, and the optimality conditions will be gradually met with theoretical guarantee. In Refs. [11] and [12], besides the states and the controls, the costates are also employed in developing the EPDE (also named the first evolution equation), and this increases the complexity of the computation. In Ref. [13], a compact version of the VEM that uses only the original variables is proposed. The costate-free optimality conditions are established and the second evolution equation is derived for the OCPs with free terminal states. In Refs. [14] and [15], the second evolution equation is furthered developed to address the OCPs with terminal equality and inequality constraints, and it succeeds to solve the general state- and\or control-constrained OCPs with path constraint in Ref. [16]. Normally, under the frame of the compact VEM, the definite conditions for the second evolution equation are required to be feasible solutions. In Refs. [17] and [18], the modified evolution equation that is valid even in the infeasible solution domain is proposed to facilitate the computation of the OCPs.

Reducing the number of variables to be sought is helpful to relieve the computation burden. However, even for the second evolution equation, both the state and the control variables will be solved along the virtual variation time dimension. In this paper, we will present the third evolution equation, which only needs to solve the control variables to get the optimal solution. Throughout the paper, our work is built upon the assumption that the solution for the optimization problem exists. We do not describe the existing conditions for the purpose of brevity. Relevant researches such as the Filippov-Cesari theorem are documented in Ref. [21]. In the following, first the principle of the VEM is briefly stated. Then the first and the second evolution equations, both derived under the variation evolution principle but specifically different, are reviewed. After that, the third evolution equation that also stems from the compact VEM is presented. Later illustrative examples are solved to test its effectiveness.

## II. PRINCIPLE OF VEM

The VEM is a newly developed method for the optimal solutions. It is enlightened from the inverse consideration of the Lyapunov dynamics stability theory in the control field [22]. As the start point of this method, the generalized Lyapunov principle for the infinite-dimensional stable continuous-time dynamics may be stated as

**Lemma 1**: For an infinite-dimensional dynamic system described by

$$\frac{\delta \boldsymbol{y}(x)}{\delta t} = \boldsymbol{f}(\boldsymbol{y}, x) \tag{1}$$

or presented equivalently in the Partial Differential Equation (PDE) form as

$$\frac{\partial \boldsymbol{y}(x,t)}{\partial t} = \boldsymbol{f}(\boldsymbol{y}, x) \tag{2}$$

where "$\delta$" denotes the variation operator and "$\partial$" denotes the partial differential operator. $t$ is the time. $x \in \mathbb{R}$ is the independent variable, $\boldsymbol{y}(x) \in \mathbb{R}^n(x)$ is the function vector of $x$, and $\boldsymbol{f} : \mathbb{R}^n(x) \times \mathbb{R} \to \mathbb{R}^n(x)$ is a vector function. Let $\hat{\boldsymbol{y}}(x)$, contained within a certain function set $\mathbb{D}(x)$, is an equilibrium function that satisfies $\boldsymbol{f}(\hat{\boldsymbol{y}}(x), x) = \boldsymbol{0}$. If there exists a continuously differentiable functional $V : \mathbb{D}(x) \to \mathbb{R}$ such that

i) $V\left(\hat{\boldsymbol{y}}(x)\right) = c$ and $V\left(\boldsymbol{y}(x)\right) > c$ in $\mathbb{D}(x) / \{\hat{\boldsymbol{y}}(x)\}$.

ii) $\frac{\delta}{\delta t}V\left(\boldsymbol{y}(x)\right)\leq 0$ in $\mathbb{D}(x)$ and $\frac{\delta}{\delta t}V\left(\boldsymbol{y}(x)\right)<0$ in $\mathbb{D}(x)/\{\hat{\boldsymbol{y}}(x)\}$.

where $c$ is a constant. Then $\boldsymbol{y}(x)=\hat{\boldsymbol{y}}(x)$ is an asymptotically stable solution in $\mathbb{D}(x)$.

The VEM analogizes the optimal solution to the asymptotically stable equilibrium point of an infinite-dimensional dynamic system, and derives such dynamics that minimize the specific performance index (act the Lyapunov functional) within an optimization problem. To implement the idea, a virtual dimension, the variation time $\tau$, is introduced to describe the process that a variable like $\boldsymbol{u}(t)$ evolves to the optimal solution to minimize the performance index under the dynamics governed by the variation dynamic evolution equations (in the form of Eq. (1)). Fig. 1 illustrates the variation evolution of control variables in the VEM to solve the OCP. Through the variation motion, the initial guess of control will evolve to the optimal solution.

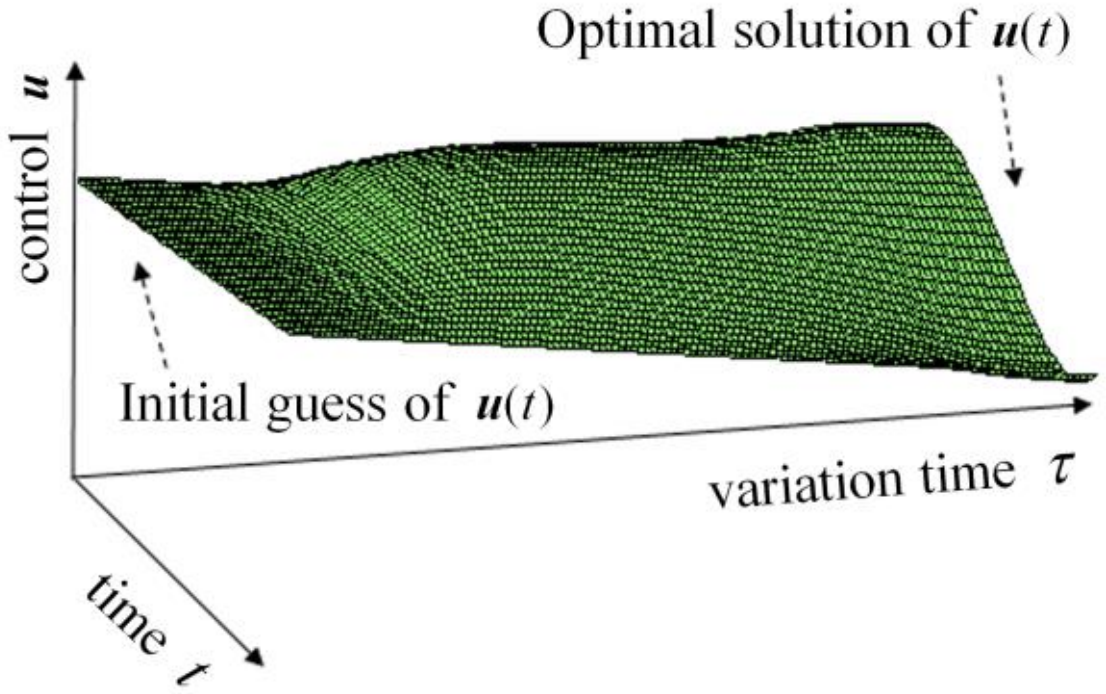

Fig. 1. The illustration of the control variable evolving along the variation time $\tau$ in the VEM.

The VEM is demonstrated for the unconstrained calculus-of-variations problems first [11][13]. The variation dynamic evolution equations, derived under the frame of the VEM, may be reformulated as the EPDE and the Evolution Differential Equation (EDE), by replacing the variation operator "$\delta$" with the partial differential operator "$\partial$" and the differential operator "d". Under the dynamics governed by the EPDE and the EDE, the variables will achieve the optimality conditions gradually. For example, consider the calculus-of-variations problems defined as

$$J=\int_{t_0}^{t_f}F\left(\boldsymbol{y}(t),\dot{\boldsymbol{y}}(t),t\right)\mathrm{d}t \tag{3}$$

where the elements of the variable vector $\boldsymbol{y}(t)\in\mathbb{R}^n$ belong to $C^2[t_0,t_f]$. $t_0$ and $t_f$ are the fixed initial and terminal time. The variation dynamic evolution equations derived are

$$\frac{\delta\boldsymbol{y}}{\delta\tau}=-\boldsymbol{K}\left(F_{\boldsymbol{y}}-\frac{\mathrm{d}}{\mathrm{d}t}(F_{\dot{\boldsymbol{y}}})\right)\ ,t\in[t_0,t_f] \tag{4}$$

$$\frac{\delta\boldsymbol{y}(t_0)}{\delta\tau}=\boldsymbol{K}\,F_{\dot{\boldsymbol{y}}}\Big|_{t_0} \tag{5}$$

$$\frac{\delta\boldsymbol{y}(t_f)}{\delta\tau}=-\boldsymbol{K}\,F_{\dot{\boldsymbol{y}}}\Big|_{t_f} \tag{6}$$

where the column vectors $F_{\boldsymbol{y}}=\frac{\partial F}{\partial\boldsymbol{y}}$ and $F_{\dot{\boldsymbol{y}}}=\frac{\partial F}{\partial\dot{\boldsymbol{y}}}$ are the shorthand notations of partial derivatives, and $\boldsymbol{K}$ is a $n\times n$ dimensional positive-definite gain matrix. Correspondingly, the reformulated EPDE and EDEs are

$$\frac{\partial \boldsymbol{y}(t,\tau)}{\partial \tau} = -\boldsymbol{K}\left( F_{\boldsymbol{y}} - \frac{\partial}{\partial t}(F_{\dot{\boldsymbol{y}}}) \right) \ , t \in [t_0, t_f] \tag{7}$$

$$\frac{\mathrm{d}\boldsymbol{y}(t_0)}{\mathrm{d}\tau} = \boldsymbol{K}\, F_{\dot{\boldsymbol{y}}}\Big|_{t_0} \tag{8}$$

$$\frac{\mathrm{d}\boldsymbol{y}(t_f)}{\mathrm{d}\tau} = -\boldsymbol{K}\, F_{\dot{\boldsymbol{y}}}\Big|_{t_f} \tag{9}$$

The equilibrium solution of Eqs. (7)-(9) will satisfy the optimality conditions, i.e., the Euler-Lagrange equation [23][24]

$$F_{\boldsymbol{y}} - \frac{\mathrm{d}}{\mathrm{d}t}(F_{\dot{\boldsymbol{y}}}) = \boldsymbol{0} \tag{10}$$

and the transversality conditions

$$F_{\dot{\boldsymbol{y}}}\Big|_{t_0} = \boldsymbol{0} \tag{11}$$

$$F_{\dot{\boldsymbol{y}}}\Big|_{t_f} = \boldsymbol{0} \tag{12}$$

With the principle of the VEM, but starting from different functional, the first and second evolution equations that solve typical OCPs are established. Since the right functions of the derived EPDEs only depend on the time $t$, they are suitable to be solved with the well-known semi-discrete method in the field of PDE numerical calculation [25]. Through the discretization along the normal time dimension, those equations are transformed to the finite-dimensional Initial-value Problems (IVPs) to be solved, with the common Ordinary Differential Equation (ODE) integration methods. Note that the resulting IVPs are defined with respect to the variation time $\tau$, not the normal time $t$.

## III. EVOLUTION EQUATIONS FOR OCP

### A. *Problem definition*

In this paper, we consider the OCPs with terminal constraint that are defined as

**Problem 1**: Consider performance index of Bolza form

$$J = \varphi(\boldsymbol{x}(t_f), t_f) + \int_{t_0}^{t_f} L\big(\boldsymbol{x}(t), \boldsymbol{u}(t), t\big)\,\mathrm{d}t \tag{13}$$

subject to the dynamic equation

$$\dot{\boldsymbol{x}} = \boldsymbol{f}(\boldsymbol{x}, \boldsymbol{u}, t) \tag{14}$$

where $t \in \mathbb{R}$ is the time. $\boldsymbol{x} \in \mathbb{R}^n$ is the state vector and its elements belong to $C^2[t_0, t_f]$. $\boldsymbol{u} \in \mathbb{R}^m$ is the control vector and its elements belong to $C^1[t_0, t_f]$. The function $L: \mathbb{R}^n \times \mathbb{R}^m \times \mathbb{R} \to \mathbb{R}$ and its first-order partial derivatives are continuous with respect to $\boldsymbol{x}$, $\boldsymbol{u}$ and $t$. The function $\varphi: \mathbb{R}^n \times \mathbb{R} \to \mathbb{R}$ and its first-order and second-order partial derivatives are continuous with respect to $\boldsymbol{x}$ and $t$. The vector function $\boldsymbol{f}: \mathbb{R}^n \times \mathbb{R}^m \times \mathbb{R} \to \mathbb{R}^n$ and its first-order partial derivatives are continuous and Lipschitz in $\boldsymbol{x}$, $\boldsymbol{u}$ and $t$. The initial time $t_0$ is fixed and the terminal time $t_f$ is free. The initial and terminal boundary conditions are respectively prescribed as

$$\boldsymbol{x}(t_0) = \boldsymbol{x}_0 \tag{15}$$

$$\boldsymbol{g}\big(\boldsymbol{x}(t_f), t_f\big) = \boldsymbol{0} \tag{16}$$

where $\boldsymbol{g}:\mathbb{R}^n \times \mathbb{R} \to \mathbb{R}^q$ is a $q$ dimensional vector function with continuous first-order partial derivatives. Find the optimal solution $(\hat{\boldsymbol{x}},\hat{\boldsymbol{u}})$ that minimizes $J$ , i.e.

$$(\hat{\boldsymbol{x}},\hat{\boldsymbol{u}}) = \arg\min(J) \tag{17}$$

In the following, we will review the first and the second evolution equations first, and then give the third evolution equation. Note that for the OCPs with fixed terminal time $t_f$ , these equations are also applicable, by ignoring terms relevant to the variation of $t_f$ . If there is no terminal constraint in the OCPs, they are just the degraded cases of Problem 1 with $\boldsymbol{g}$ vanishing.

*B. The first evolution equation*

The first evolution equation that solves Problem 1 are derived through a constructed unconstrained functional, which employs the classic optimality conditions as

$$\begin{aligned}\overline{J} = &\left(\boldsymbol{x}(t_0)-\boldsymbol{x}_0\right)^{\mathrm{T}} \boldsymbol{W}_{\boldsymbol{x}_0}\left(\boldsymbol{x}(t_0)-\boldsymbol{x}_0\right)+\boldsymbol{g}^{\mathrm{T}}\boldsymbol{W}_{\boldsymbol{x}_f}\boldsymbol{g}+\left(\boldsymbol{\lambda}(t_f)-\varphi_{\boldsymbol{x}}(t_f)-\boldsymbol{g}_{\boldsymbol{x}}^{\mathrm{T}}(t_f)\boldsymbol{\pi}\right)^{\mathrm{T}}\boldsymbol{W}_{\boldsymbol{\lambda}_f}\left(\boldsymbol{\lambda}(t_f)-\varphi_{\boldsymbol{x}}(t_f)-\boldsymbol{g}_{\boldsymbol{x}}^{\mathrm{T}}(t_f)\boldsymbol{\pi}\right)\\ &+\left(H(t_f)+\varphi_t(t_f)+\boldsymbol{\pi}^{\mathrm{T}}\boldsymbol{g}_t(t_f)\right)^2+\int_{t_0}^{t_f}\left\{\left(\dot{\boldsymbol{x}}-H_{\boldsymbol{\lambda}}\right)^{\mathrm{T}}\left(\dot{\boldsymbol{x}}-H_{\boldsymbol{\lambda}}\right)+\left(\dot{\boldsymbol{\lambda}}+H_{\boldsymbol{x}}\right)^{\mathrm{T}}\left(\dot{\boldsymbol{\lambda}}+H_{\boldsymbol{x}}\right)+H_{\boldsymbol{u}}^{\mathrm{T}}H_{\boldsymbol{u}}\right\}\mathrm{d}t\end{aligned} \tag{18}$$

where $H = L+\boldsymbol{\lambda}^{\mathrm{T}}\boldsymbol{f}$ is the Hamiltonian, $\boldsymbol{\lambda}\in\mathbb{R}^n$ is the costates, and $\boldsymbol{\pi}\in\mathbb{R}^q$ is the Lagrange multipliers that adjoin the terminal constraint (16) in the classic adjoining method [26]. $\boldsymbol{W}_{\boldsymbol{x}_0}$ , $\boldsymbol{W}_{\boldsymbol{x}_f}$ and $\boldsymbol{W}_{\boldsymbol{\lambda}_f}$ are right dimensional weighted positive-definite matrixes. "$\mathrm{T}$" is the transpose operator. Obviously, the minimum solution of the unconstrained functional (18) satisfies the feasibility conditions (14)-(16) and the classic optimality conditions, i.e.

$$\dot{\boldsymbol{\lambda}}+H_{\boldsymbol{x}} = \dot{\boldsymbol{\lambda}}+L_{\boldsymbol{x}}+\boldsymbol{f}_{\boldsymbol{x}}^{\mathrm{T}}\boldsymbol{\lambda}=\boldsymbol{0} \tag{19}$$

$$H_{\boldsymbol{u}} = L_{\boldsymbol{u}}+\boldsymbol{f}_{\boldsymbol{u}}^{\mathrm{T}}\boldsymbol{\lambda}=\boldsymbol{0} \tag{20}$$

and the transversality conditions

$$H(t_f)+\varphi_t(t_f)+\boldsymbol{\pi}^{\mathrm{T}}\boldsymbol{g}_t(t_f)=0 \tag{21}$$

$$\boldsymbol{\lambda}(t_f)-\varphi_{\boldsymbol{x}}(t_f)-\boldsymbol{g}_{\boldsymbol{x}}^{\mathrm{T}}(t_f)\boldsymbol{\pi}=\boldsymbol{0} \tag{22}$$

With the VEM, the EPDE and the EDEs derived from the unconstrained functional (18) (regarded as the Lyapunov functional) are [11]

$$\frac{\partial}{\partial\tau}\begin{bmatrix}\boldsymbol{x}(t,\tau)\\ \boldsymbol{\lambda}(t,\tau)\\ \boldsymbol{u}(t,\tau)\end{bmatrix}=-\boldsymbol{K}\boldsymbol{z}, \quad t\in(t_0,t_f) \tag{23}$$

$$\frac{\mathrm{d}}{\mathrm{d}\tau}\begin{bmatrix}\boldsymbol{x}(t_0)\\ \boldsymbol{\lambda}(t_0)\\ \boldsymbol{u}(t_0)\end{bmatrix}=\boldsymbol{K}\left.\begin{bmatrix}\boldsymbol{W}_{\boldsymbol{x}_0}\left(\boldsymbol{x}_0-\boldsymbol{x}\right)+\left(\dot{\boldsymbol{x}}-H_{\boldsymbol{\lambda}}\right)\\ \left(\dot{\boldsymbol{\lambda}}+H_{\boldsymbol{x}}\right)\\ \boldsymbol{z}_{\boldsymbol{u}}\end{bmatrix}\right\|_{t_0} \tag{24}$$

$$\frac{\mathrm{d}}{\mathrm{d}\tau}\begin{bmatrix}\boldsymbol{x}(t_f)\\ \boldsymbol{\lambda}(t_f)\\ \boldsymbol{u}(t_f)\end{bmatrix}=-\boldsymbol{K}\left.\begin{bmatrix}\boldsymbol{g}_{\boldsymbol{x}}^{\mathrm{T}}\boldsymbol{W}_{\boldsymbol{x}_f}\boldsymbol{g}-\left(\varphi_{\boldsymbol{x}\boldsymbol{x}}+\dfrac{\partial(\boldsymbol{g}_{\boldsymbol{x}}^{\mathrm{T}}\boldsymbol{\pi})}{\partial\boldsymbol{x}}\right)^{\mathrm{T}}\boldsymbol{W}_{\boldsymbol{\lambda}_f}\left(\boldsymbol{\lambda}-\varphi_{\boldsymbol{x}}-\boldsymbol{g}_{\boldsymbol{x}}^{\mathrm{T}}\boldsymbol{\pi}\right)+(H+\varphi_t+\boldsymbol{\pi}^{\mathrm{T}}\boldsymbol{g}_t)(H_{\boldsymbol{x}}+\varphi_{\boldsymbol{x}t}+\boldsymbol{\pi}^{\mathrm{T}}\boldsymbol{g}_{\boldsymbol{x}t})+\left(\dot{\boldsymbol{x}}-H_{\boldsymbol{\lambda}}\right)\\ \boldsymbol{W}_{\boldsymbol{\lambda}_f}\left(\boldsymbol{\lambda}-\varphi_{\boldsymbol{x}}-\boldsymbol{g}_{\boldsymbol{x}}^{\mathrm{T}}\boldsymbol{\pi}\right)+\left(\dot{\boldsymbol{\lambda}}+H_{\boldsymbol{x}}\right)\\ \left(H+\varphi_t+\boldsymbol{\pi}^{\mathrm{T}}\boldsymbol{g}_t\right)H_{\boldsymbol{u}}\end{bmatrix}\right\|_{t_f} \tag{25}$$

$$\frac{\mathrm{d}t_f}{\mathrm{d}\tau} = -k_{t_f} h(t_f) \tag{26}$$

$$\frac{\mathrm{d}\boldsymbol{\pi}}{\mathrm{d}\tau} = \boldsymbol{K}_{\boldsymbol{\pi}} \boldsymbol{g}_{\boldsymbol{x}}(t_f) \boldsymbol{W}_{\lambda_f} \left( \boldsymbol{\lambda}(t_f) - \varphi_{\boldsymbol{x}}(t_f) - \boldsymbol{g}_{\boldsymbol{x}}^{\mathrm{T}}(t_f) \boldsymbol{\pi} \right) \tag{27}$$

where

$$z(t) = \begin{bmatrix} z_{\boldsymbol{x}} \\ z_{\lambda} \\ z_{\boldsymbol{u}} \end{bmatrix} = H_{\boldsymbol{yy}} \begin{bmatrix} \left( H_{\boldsymbol{x}} + \dfrac{\partial \boldsymbol{\lambda}}{\partial t} \right) \\ \left( \boldsymbol{f} - \dfrac{\partial \boldsymbol{x}}{\partial t} \right) \\ H_{\boldsymbol{u}} \end{bmatrix} - \frac{\partial}{\partial t} \begin{bmatrix} \left( \dfrac{\partial \boldsymbol{x}}{\partial t} - \boldsymbol{f} \right) \\ \left( \dfrac{\partial \boldsymbol{\lambda}}{\partial t} + H_{\boldsymbol{x}} \right) \\ \boldsymbol{0} \end{bmatrix} \tag{28}$$

$$\begin{aligned} h = {} & 2\boldsymbol{g}_t^{\mathrm{T}} \boldsymbol{W}_{\boldsymbol{x}_f} \boldsymbol{g} + 2(H + \varphi_t + \boldsymbol{\pi}^{\mathrm{T}} \boldsymbol{g}_t)(H_t + \varphi_{tt} + \boldsymbol{\pi}^{\mathrm{T}} \boldsymbol{g}_{tt}) \\ & - \left(\frac{\partial \boldsymbol{x}(t,\tau)}{\partial t}\right)^{\mathrm{T}} \frac{\partial \boldsymbol{x}(t,\tau)}{\partial t} - \left(\frac{\partial \boldsymbol{\lambda}(t,\tau)}{\partial t}\right)^{\mathrm{T}} \frac{\partial \boldsymbol{\lambda}(t,\tau)}{\partial t} + H_{\boldsymbol{x}}^{\mathrm{T}} H_{\boldsymbol{x}} + \boldsymbol{f}^{\mathrm{T}} \boldsymbol{f} + H_{\boldsymbol{u}}^{\mathrm{T}} H_{\boldsymbol{u}} \end{aligned} \tag{29}$$

$H_{\boldsymbol{yy}} = \begin{bmatrix} H_{\boldsymbol{xx}} & \boldsymbol{f}_{\boldsymbol{x}}^{\mathrm{T}} & H_{\boldsymbol{xu}} \\ \boldsymbol{f}_{\boldsymbol{x}} & \boldsymbol{0} & \boldsymbol{f}_{\boldsymbol{u}} \\ H_{\boldsymbol{ux}} & \boldsymbol{f}_{\boldsymbol{u}}^{\mathrm{T}} & H_{\boldsymbol{uu}} \end{bmatrix}$ is the Hessian matrix of $H$, and $\boldsymbol{f}_{\boldsymbol{x}}$ and $\boldsymbol{f}_{\boldsymbol{u}}$ are the Jacobi matrixes of the dynamics $\boldsymbol{f}$. $\boldsymbol{g}_t$ and $\boldsymbol{g}_{\boldsymbol{x}}$ are the first-order partial derivatives, and $\boldsymbol{g}_{tt}$ and $\boldsymbol{g}_{\boldsymbol{x}t}$ are second-order partial derivatives of the function $\boldsymbol{g}$. $\varphi_t$ and $\varphi_{\boldsymbol{x}}$ are the first-order partial derivatives, and $\varphi_{\boldsymbol{x}t}$ and $\varphi_{\boldsymbol{xx}}$ are second-order partial derivatives of the function $\varphi$. $\boldsymbol{K}$ is a $(2n+m)\times(2n+m)$ dimensional positive-definite gain matrix, $k_{t_f}$ is a positive gain constant, and $\boldsymbol{K}_{\boldsymbol{\pi}}$ is a $q\times q$ dimensional positive-definite gain matrix. For the evolution equations (23)-(27), the definite conditions including $\left.\begin{bmatrix} \boldsymbol{x}(t,\tau) \\ \boldsymbol{\lambda}(t,\tau) \\ \boldsymbol{u}(t,\tau) \end{bmatrix}\right|_{\tau=0} = \begin{bmatrix} \tilde{\boldsymbol{x}}(t) \\ \tilde{\boldsymbol{\lambda}}(t) \\ \tilde{\boldsymbol{u}}(t) \end{bmatrix}$, $t_f\big|_{\tau=0} = \tilde{t}_f$, and $\boldsymbol{\pi}\big|_{\tau=0} = \tilde{\boldsymbol{\pi}}$ may be arbitrary solutions, and the evolving solutions will gradually meet the feasibility conditions (14)-(16) and the optimality conditions (19)-(22).

Employing the first evolution equation to get the optimal solution, the anticipated variable evolving along the variation time $\tau$, as depicted in Fig. 1, includes the original (state and control) variables and augmented costate variables. When we apply the semi-discrete method to solve the EPDE (23), the state, the costate, and the control variables all need to be discretized along the normal time dimension $t$.

### *C. The second evolution equation*

Through regarding the primary performance index (13) as the Lyapunov functional, the second evolution equations derived in the feasible solution domain $\mathbb{D}_o$ (in which any solution satisfies Eqs. (14)-(16)) are [14]

$$\frac{\partial \boldsymbol{x}(t,\tau)}{\partial \tau} = \int_{t_0}^{t} \boldsymbol{\Phi}_o(t,s) \boldsymbol{f}_{\boldsymbol{u}}(s) \frac{\partial \boldsymbol{u}(s,\tau)}{\partial \tau} \mathrm{d}s \tag{30}$$

$$\frac{\partial \boldsymbol{u}(t,\tau)}{\partial \tau} = -\boldsymbol{K} \left( \boldsymbol{p}_{\boldsymbol{u}} + \boldsymbol{f}_{\boldsymbol{u}}^{\mathrm{T}} \boldsymbol{\Phi}_o^{\mathrm{T}}(t_f, t) \boldsymbol{g}_{\boldsymbol{x}}^{\mathrm{T}}(t_f) \boldsymbol{\pi} \right) \tag{31}$$

$$\frac{\mathrm{d}t_f}{\mathrm{d}\tau} = -k_{t_f} \left( L + \varphi_t + \varphi_{\boldsymbol{x}}^{\mathrm{T}} \boldsymbol{f} + \boldsymbol{\pi}^{\mathrm{T}} (\boldsymbol{g}_{\boldsymbol{x}} \boldsymbol{f} + \boldsymbol{g}_t) \right)\Big|_{t_f} \tag{32}$$

where

$$\boldsymbol{p}_{\boldsymbol{u}}(t)=L_{\boldsymbol{u}}+\boldsymbol{f}_{\boldsymbol{u}}{}^{\mathrm{T}}\varphi_{\boldsymbol{x}}+\boldsymbol{f}_{\boldsymbol{u}}{}^{\mathrm{T}}\left(\int_{t}^{t_f}\boldsymbol{\Phi}_o{}^{\mathrm{T}}(\sigma,t)\Big(L_{\boldsymbol{x}}(\sigma)+\varphi_{t\boldsymbol{x}}(\sigma)+\varphi_{\boldsymbol{xx}}{}^{\mathrm{T}}(\sigma)\boldsymbol{f}(\sigma)+\boldsymbol{f}_{\boldsymbol{x}}{}^{\mathrm{T}}(\sigma)\varphi_{\boldsymbol{x}}(\sigma)\Big)\mathrm{d}\sigma\right) \tag{33}$$

$L_{\boldsymbol{x}}$ and $L_{\boldsymbol{u}}$ are the partial derivatives of $L$ with respect to $\boldsymbol{x}$ and $\boldsymbol{u}$. $\boldsymbol{\Phi}_o(t,s)$ is the $n\times n$ dimensional state transition matrix from time point $s$ to time point $t$, which satisfies

$$\frac{\partial}{\partial t}\boldsymbol{\Phi}_o(t,s)=\boldsymbol{f}_{\boldsymbol{x}}(t)\boldsymbol{\Phi}_o(t,s) \tag{34}$$

The parameter vector $\boldsymbol{\pi}\in\mathbb{R}^q$ is calculated by

$$\boldsymbol{\pi}=-\boldsymbol{M}^{-1}\boldsymbol{r} \tag{35}$$

where the $q\times q$ dimensional matrix $\boldsymbol{M}$ and the $q$ dimensional vector $\boldsymbol{r}$ are

$$\boldsymbol{M}=\boldsymbol{g}_{\boldsymbol{x}}(t_f)\left(\int_{t_0}^{t_f}\boldsymbol{\Phi}_o(t_f,t)\boldsymbol{f}_{\boldsymbol{u}}\boldsymbol{K}\boldsymbol{f}_{\boldsymbol{u}}{}^{\mathrm{T}}\boldsymbol{\Phi}_o{}^{\mathrm{T}}(t_f,t)\,\mathrm{d}t\right)\boldsymbol{g}_{\boldsymbol{x}}{}^{\mathrm{T}}(t_f)+k_{t_f}\left\{(\boldsymbol{g}_{\boldsymbol{x}}\boldsymbol{f}+\boldsymbol{g}_t)(\boldsymbol{g}_{\boldsymbol{x}}\boldsymbol{f}+\boldsymbol{g}_t)^{\mathrm{T}}\right\}\Big|_{t_f} \tag{36}$$

$$\boldsymbol{r}=\boldsymbol{g}_{\boldsymbol{x}}(t_f)\left(\int_{t_0}^{t_f}\boldsymbol{\Phi}_o(t_f,t)\boldsymbol{f}_{\boldsymbol{u}}\boldsymbol{K}\boldsymbol{p}_{\boldsymbol{u}}\,\mathrm{d}t\right)+k_{t_f}\left\{(\boldsymbol{g}_{\boldsymbol{x}}\boldsymbol{f}+\boldsymbol{g}_t)(\varphi_t+\varphi_{\boldsymbol{x}}{}^{\mathrm{T}}\boldsymbol{f}+L)\right\}\Big|_{t_f} \tag{37}$$

$\boldsymbol{K}$ is the $m\times m$ dimensional positive-definite gain matrix and $k_{t_f}$ is a positive gain constant. For Eqs. (30)-(32), the definite conditions are $\begin{bmatrix}\boldsymbol{x}(t,\tau)\\ \boldsymbol{u}(t,\tau)\end{bmatrix}\Bigg|_{\tau=0}=\begin{bmatrix}\tilde{\boldsymbol{x}}(t)\\ \tilde{\boldsymbol{u}}(t)\end{bmatrix}$ and $t_f\big|_{\tau=0}=\tilde{t}_f$, where $\tilde{\boldsymbol{x}}(t)$ and $\tilde{\boldsymbol{u}}(t)$ are the feasible initial solutions with terminal time $\tilde{t}_f$. Use this formulation, the solution of the evolution equations maintains the feasibility conditions (14)-(16) and it will satisfy the costate-free optimality conditions at $\tau=+\infty$, i.e.

$$\boldsymbol{p}_{\boldsymbol{u}}+\boldsymbol{f}_{\boldsymbol{u}}{}^{\mathrm{T}}\boldsymbol{\Phi}_o{}^{\mathrm{T}}(t_f,t)\boldsymbol{g}_{\boldsymbol{x}}{}^{\mathrm{T}}(t_f)\boldsymbol{\pi}=\boldsymbol{0} \tag{38}$$

$$L+\varphi_t+\varphi_{\boldsymbol{x}}{}^{\mathrm{T}}\boldsymbol{f}+\boldsymbol{\pi}^{\mathrm{T}}(\boldsymbol{g}_{\boldsymbol{x}}\boldsymbol{f}+\boldsymbol{g}_t)\Big|_{t_f}=0 \tag{39}$$

With the first-order stable dynamics principle, Eqs. (30)-(32) are further modified to be valid even in the infeasible solution domain as (See Ref. [18])

$$\frac{\partial\boldsymbol{x}(t,\tau)}{\partial\tau}=-\boldsymbol{\Phi}_o(t,t_0)\boldsymbol{K}_{\boldsymbol{x}_0}\left(\boldsymbol{x}(t_0)-\boldsymbol{x}_0\right)+\int_{t_0}^{t}\boldsymbol{\Phi}_o(t,s)\boldsymbol{f}_{\boldsymbol{u}}(s)\frac{\partial\boldsymbol{u}(s,\tau)}{\partial\tau}\mathrm{d}s-\int_{t_0}^{t}\boldsymbol{\Phi}_o(t,s)\boldsymbol{K}_f\left(\frac{\partial\boldsymbol{x}(s,\tau)}{\partial s}-\boldsymbol{f}(s)\right)\mathrm{d}s \tag{40}$$

$$\frac{\partial\boldsymbol{u}(t,\tau)}{\partial\tau}=-\boldsymbol{K}\left(\overline{\boldsymbol{p}}_{\boldsymbol{u}}+\boldsymbol{f}_{\boldsymbol{u}}{}^{\mathrm{T}}\boldsymbol{\Phi}_o{}^{\mathrm{T}}(t_f,t)\boldsymbol{g}_{\boldsymbol{x}}{}^{\mathrm{T}}(t_f)\boldsymbol{\pi}\right) \tag{41}$$

$$\frac{\mathrm{d}t_f}{\mathrm{d}\tau}=-k_{t_f}\left(L+\varphi_t+\varphi_{\boldsymbol{x}}{}^{\mathrm{T}}\frac{\partial\boldsymbol{x}(t,\tau)}{\partial t}+\boldsymbol{\pi}^{\mathrm{T}}(\boldsymbol{g}_{\boldsymbol{x}}\frac{\partial\boldsymbol{x}(t,\tau)}{\partial t}+\boldsymbol{g}_t)\right)\Bigg|_{t_f} \tag{42}$$

where

$$\overline{\boldsymbol{p}}_u(t)=L_{\boldsymbol{u}}+\boldsymbol{f}_{\boldsymbol{u}}{}^{\mathrm{T}}\varphi_{\boldsymbol{x}}+\boldsymbol{f}_{\boldsymbol{u}}{}^{\mathrm{T}}\left(\int_{t}^{t_f}\boldsymbol{\Phi}_o{}^{\mathrm{T}}(\sigma,t)\left(L_{\boldsymbol{x}}(\sigma)+\varphi_{t\boldsymbol{x}}(\sigma)+\varphi_{\boldsymbol{xx}}{}^{\mathrm{T}}(\sigma)\frac{\partial\boldsymbol{x}(\sigma,\tau)}{\partial\sigma}+\boldsymbol{f}_{\boldsymbol{x}}{}^{\mathrm{T}}(\sigma)\varphi_{\boldsymbol{x}}(\sigma)\right)\mathrm{d}\sigma\right) \tag{43}$$

$\boldsymbol{\pi}$ is computed by Eq. (35) while $\boldsymbol{M}$ and $\boldsymbol{r}$ are modified as

$$\boldsymbol{M}=\boldsymbol{g}_{\boldsymbol{x}}(t_f)\left(\int_{t_0}^{t_f}\boldsymbol{\Phi}_o(t_f,t)\boldsymbol{f}_{\boldsymbol{u}}\boldsymbol{K}\boldsymbol{f}_{\boldsymbol{u}}{}^{\mathrm{T}}\boldsymbol{\Phi}_o{}^{\mathrm{T}}(t_f,t)\,\mathrm{d}t\right)\boldsymbol{g}_{\boldsymbol{x}}{}^{\mathrm{T}}(t_f)+k_{t_f}\left\{(\boldsymbol{g}_{\boldsymbol{x}}\frac{\partial\boldsymbol{x}(t,\tau)}{\partial t}+\boldsymbol{g}_t)(\boldsymbol{g}_{\boldsymbol{x}}\frac{\partial\boldsymbol{x}(t,\tau)}{\partial t}+\boldsymbol{g}_t)^{\mathrm{T}}\right\}\Bigg|_{t_f} \tag{44}$$

$$\begin{aligned}\boldsymbol{r} = {} & \boldsymbol{g}_{\boldsymbol{x}}(t_f)\left(\int_{t_0}^{t_f}\boldsymbol{\Phi}_o(t_f,t)\boldsymbol{f}_{\boldsymbol{u}}\boldsymbol{K}\boldsymbol{p}_{\boldsymbol{u}}\,\mathrm{d}t\right)+k_{t_f}\left.\left\{(\boldsymbol{g}_{\boldsymbol{x}}\frac{\partial \boldsymbol{x}(t,\tau)}{\partial t}+\boldsymbol{g}_t)(\varphi_t+{\varphi_{\boldsymbol{x}}}^{\mathrm{T}}\frac{\partial \boldsymbol{x}(t,\tau)}{\partial t}+L)\right\}\right|_{t_f} \\ & -\boldsymbol{K}_{\boldsymbol{g}}\boldsymbol{g}+\boldsymbol{g}_{\boldsymbol{x}}(t_f)\boldsymbol{\Phi}_o(t_f,t_0)\boldsymbol{K}_{x_0}\left(\boldsymbol{x}(t_0)-\boldsymbol{x}_0\right)+\boldsymbol{g}_{\boldsymbol{x}}(t_f)\int_{t_0}^{t_f}\boldsymbol{\Phi}_o(t,s)\boldsymbol{K}_f\left(\frac{\partial \boldsymbol{x}(s,\tau)}{\partial s}-\boldsymbol{f}(s)\right)\mathrm{d}s\end{aligned} \tag{45}$$

$\boldsymbol{K}_{\boldsymbol{g}}$ is a $q\times q$ dimensional positive-definite matrix. $\boldsymbol{K}_{x_0}$ and $\boldsymbol{K}_{\boldsymbol{f}}$ are $n\times n$ dimensional positive-definite matrixes. For the modified formulation, the definite conditions $\left.\begin{bmatrix}\boldsymbol{x}(t,\tau)\\ \boldsymbol{u}(t,\tau)\end{bmatrix}\right|_{\tau=0}$ and $\left.t_f\right|_{\tau=0}$ may be arbitrary solutions, and its solution will ultimately satisfy the feasibility conditions (14)-(16) and the optimality conditions (38), (39).

On the modified evolution equations, the interesting thing is that we modify the equation regarding the states to eliminate the initial state error and the dynamics error, and modify the equation regarding the controls to eliminate the error in the terminal constraint. Here, we introduce another function set, the quasi-feasible solution domain $\bar{\mathbb{D}}_o$, in which the variables satisfy the dynamic constraint (14) and the initial boundary conditions (15). Since finding a solution in $\bar{\mathbb{D}}_o$ as the definite condition is easy, simply by numerical integration, then Eqs. (40)-(42) return back to the forms of Eqs. (30)-(32) when operating in $\bar{\mathbb{D}}_o$. Note that now $\boldsymbol{r}$ in Eq. (45) that determines $\boldsymbol{\pi}$ is modified as

$$\boldsymbol{r} = \boldsymbol{g}_{\boldsymbol{x}}(t_f)\left(\int_{t_0}^{t_f}\boldsymbol{\Phi}_o(t_f,t)\boldsymbol{f}_{\boldsymbol{u}}\boldsymbol{K}\boldsymbol{p}_{\boldsymbol{u}}\,\mathrm{d}t\right)+k_{t_f}\left.\left\{(\boldsymbol{g}_{\boldsymbol{x}}\boldsymbol{f}+\boldsymbol{g}_t)(\varphi_t+{\varphi_{\boldsymbol{x}}}^{\mathrm{T}}\boldsymbol{f}+L)\right\}\right|_{t_f}-\boldsymbol{K}_{\boldsymbol{g}}\boldsymbol{g} \tag{46}$$

and $\boldsymbol{M}$ returns to Eq. (36). For this formulation, the solution will always satisfy the dynamic constraint (14) and the initial boundary conditions (15), and it will meet the terminal constraints (16) and the optimality conditions (38), (39) gradually.

Use the second evolution equation, the anticipated variable evolving along the variation time $\tau$, as depicted in Fig. 1, includes the state variables and the control variables. In seeking the optimal solutions with the numerical method, both the state and the control variables in the EPDEs (30) and (31) need to be discretized along the normal time dimension $t$.

*D. The third evolution equation*

The third evolution equation also stems from the compact VEM. However, we do not solve for the states from their variation motion while they are computed in terms of the state equation (14). The formulation of the third evolution equation is

$$\frac{\partial \boldsymbol{x}(t,\tau)}{\partial t} = \boldsymbol{f}(\boldsymbol{x},\boldsymbol{u},t) \tag{47}$$

$$\frac{\partial \boldsymbol{u}(t,\tau)}{\partial \tau} = -\boldsymbol{K}\left(\boldsymbol{p}_{\boldsymbol{u}}+{\boldsymbol{f}_{\boldsymbol{u}}}^{\mathrm{T}}{\boldsymbol{\Phi}_o}^{\mathrm{T}}(t_f,t){\boldsymbol{g}_{\boldsymbol{x}}}^{\mathrm{T}}(t_f)\boldsymbol{\pi}\right) \tag{48}$$

$$\frac{\mathrm{d}t_f}{\mathrm{d}\tau} = -k_{t_f}\left.\left(L+\varphi_t+{\varphi_{\boldsymbol{x}}}^{\mathrm{T}}\boldsymbol{f}+\boldsymbol{\pi}^{\mathrm{T}}(\boldsymbol{g}_{\boldsymbol{x}}\boldsymbol{f}+\boldsymbol{g}_t)\right)\right|_{t_f} \tag{49}$$

where $\boldsymbol{p}_{\boldsymbol{u}}$ is defined in Eq. (33). $\boldsymbol{\pi}$ is computed by Eq. (35) with $\boldsymbol{M}$ defined in Eq. (36) and $\boldsymbol{r}$ defined in Eq. (46). $\boldsymbol{K}$ and $k_{t_f}$ are the right dimensional gain parameters. The definite conditions are $\left.\boldsymbol{x}(t,\tau)\right|_{t=0}=\boldsymbol{x}_0$, $\left.\boldsymbol{u}(t,\tau)\right|_{\tau=0}=\tilde{\boldsymbol{u}}(t)$, and $\left.t_f\right|_{\tau=0}=\tilde{t}_f$. Actually, Eq. (47) may be independently calculated (at least numerically) from its definite condition as

$$\boldsymbol{x}(t,\tau) = \boldsymbol{x}_0+\int_{t_0}^{t}\boldsymbol{f}\left(\boldsymbol{x}(s,\tau),\boldsymbol{u}(s,\tau),s\right)\mathrm{d}s \tag{50}$$

Thus only Eqs. (48) and (49) take effect in this formulation. Then the definite conditions for the third evolution equation are $\boldsymbol{u}(t,\tau)\big|_{\tau=0}$ and $t_f\big|_{\tau=0}$, and they may be arbitrary solutions. This evolution equation also guarantees that the terminal constraints (16) and the optimality conditions (38), (39) will be satisfied gradually.

With the third evolution equation, the anticipated variable evolving along the variation time $\tau$, as depicted in Fig. 1, only includes the control variables. For Eq. (48), it may be solved with the semi-discrete method as well. Then only the control variables need to be discretized for the numerical computation.

## IV. ILLUSTRATIVE EXAMPLES

First a linear example taken from Xie [27] is solved.

**Example 1**: Consider the following dynamic system

$$\dot{\boldsymbol{x}} = \boldsymbol{A}\boldsymbol{x} + \boldsymbol{b}u$$

where $\boldsymbol{x} = \begin{bmatrix} x_1 \\ x_2 \end{bmatrix}$, $\boldsymbol{A} = \begin{bmatrix} 0 & 1 \\ 0 & 0 \end{bmatrix}$, and $\boldsymbol{b} = \begin{bmatrix} 0 \\ 1 \end{bmatrix}$. Find the solution that minimizes the performance index

$$J = \frac{1}{2}\int_{t_0}^{t_f} u^2 \mathrm{d}t$$

with the boundary conditions

$$\boldsymbol{x}(t_0) = \begin{bmatrix} 1 \\ 1 \end{bmatrix}, \ \boldsymbol{x}(t_f) = \begin{bmatrix} 0 \\ 0 \end{bmatrix}$$

where the initial time $t_0 = 0$ and the terminal time $t_f = 2$ are fixed.

This example was solved with the first evolution equation in Ref. [11] and the second evolution equation in Ref. [14]. Here we solve it with the third evolution equation, whose concrete form was derived as

$$\frac{\partial u(t,\tau)}{\partial \tau} = -K\left\{u + \boldsymbol{b}^{\mathrm{T}}\left(e^{\boldsymbol{A}(t_f - t)}\right)^{\mathrm{T}}\boldsymbol{\pi}\right\}$$

where $\boldsymbol{\pi}$ was calculated by Eq. (35) with $\boldsymbol{M}$ and $\boldsymbol{r}$ being

$$\boldsymbol{M} = K\int_{t_0}^{t_f} e^{\boldsymbol{A}(t_f - t)}\boldsymbol{b}\boldsymbol{b}^{\mathrm{T}}\left(e^{\boldsymbol{A}(t_f - t)}\right)^{\mathrm{T}}\mathrm{d}t$$

$$\boldsymbol{r} = K\int_{t_0}^{t_f} e^{\boldsymbol{A}(t_f - t)}\boldsymbol{b}u\,\mathrm{d}t - \boldsymbol{K}_{\boldsymbol{g}}\boldsymbol{x}(t_f)$$

The state variables were solved by

$$\boldsymbol{x}(t,\tau) = e^{\boldsymbol{A}(t-t_0)}\begin{bmatrix} 1 \\ 1 \end{bmatrix} + \int_{t_0}^{t} e^{\boldsymbol{A}(t-s)}\boldsymbol{b}u(s,\tau)\,\mathrm{d}s$$

The one-dimensional gain matrix $K$ was set as $K = 0.1$. The $2\times2$ dimensional matrix $\boldsymbol{K}_{\boldsymbol{g}}$ was $\begin{bmatrix} 0.1 & 0 \\ 0 & 0.1 \end{bmatrix}$. The definite conditions of the PDE was $u(t,\tau)\big|_{\tau=0} = 0$. Using the semi-discrete method, the time horizon $[t_0,\ t_f]$ was discretized uniformly with 41 points. Thus, a dynamic system with 41 states was obtained and the OCP was transformed to a finite-dimensional IVP. Note that in this discretization granularity, there will be 123 states from the second evolution equation formulation and will be 205 states for the first evolution equation. Huge difference. The ODE integrator “ode45” in Matlab, with default relative error tolerance $1\times10^{-3}$ and default absolute error tolerance $1\times10^{-6}$, was employed to solve the IVP. The spline interpolation was used to obtain the control within the discretization points. For comparison, the analytic solution by solving the BVP is also presented.

$$\begin{aligned}
\hat{x}_1 &= 0.5t^3 - 1.75t^2 + t + 1 \\
\hat{x}_2 &= 1.5t^2 - 3.5t + 1 \\
\hat{\lambda}_1 &= 3 \\
\hat{\lambda}_2 &= -3t + 3.5 \\
\hat{u} &= 3t - 3.5
\end{aligned}$$

Fig. 2 shows the evolving process of $u(t)$ solutions towards the optimal. At $\tau$ = 300s, the numerical solution is indistinguishable from the optimal, and this shows the effectiveness of the third evolution equation. Fig. 3 plots the numerical solutions of $x_1(t)$ and $x_2(t)$ at $\tau$ = 300s. They are almost identical with the analytic solutions. Fig. 4 plots the profile of performance index value against the variation time. Since starting from a zero solution, its value increases from zero, and then monotonously approaches the minimum. It almost reaches the analytic value of $\hat{J}$ = 3.25 after $\tau$ =100s. In Fig. 5, the evolution profiles of the Lagrange multipliers are presented. They start from the value of $\boldsymbol{\pi}\big|_{\tau=0} = \begin{bmatrix} 2.9963 \\ -2.4963 \end{bmatrix}$, and we compute that $\boldsymbol{\pi}\big|_{\tau=300} = \begin{bmatrix} 3.0000 \\ -2.5000 \end{bmatrix}$. From the analytic relation to the costates uncovered in Ref. [16], we have

$$\boldsymbol{\lambda}(t) = \boldsymbol{\Phi}_o^{\mathrm{T}}(t_f, t)\boldsymbol{\pi} = \left(e^{A(t_f - t)}\right)^{\mathrm{T}} \boldsymbol{\pi} = \begin{bmatrix} 1 & 0 \\ t_f - t & 1 \end{bmatrix} \begin{bmatrix} 3.0000 \\ -2.5000 \end{bmatrix} = \begin{bmatrix} 3.0000 \\ -3.0000t + 3.5000 \end{bmatrix}$$

This is exactly same to the analytic solution of $\begin{bmatrix} \hat{\lambda}_1 \\ \hat{\lambda}_2 \end{bmatrix}$.

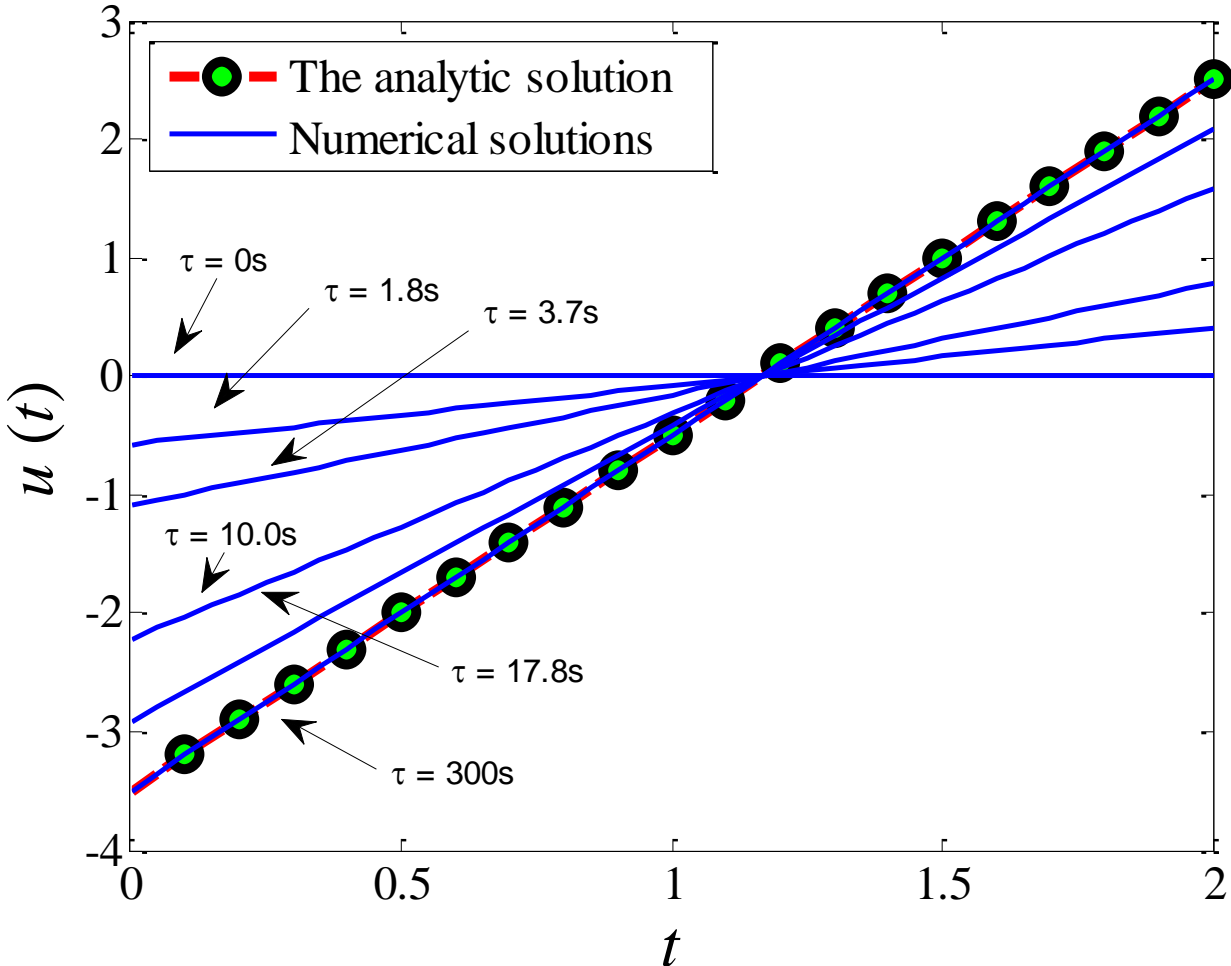

Fig. 2 The evolution of numerical solutions of $u$ to the optimal solution.

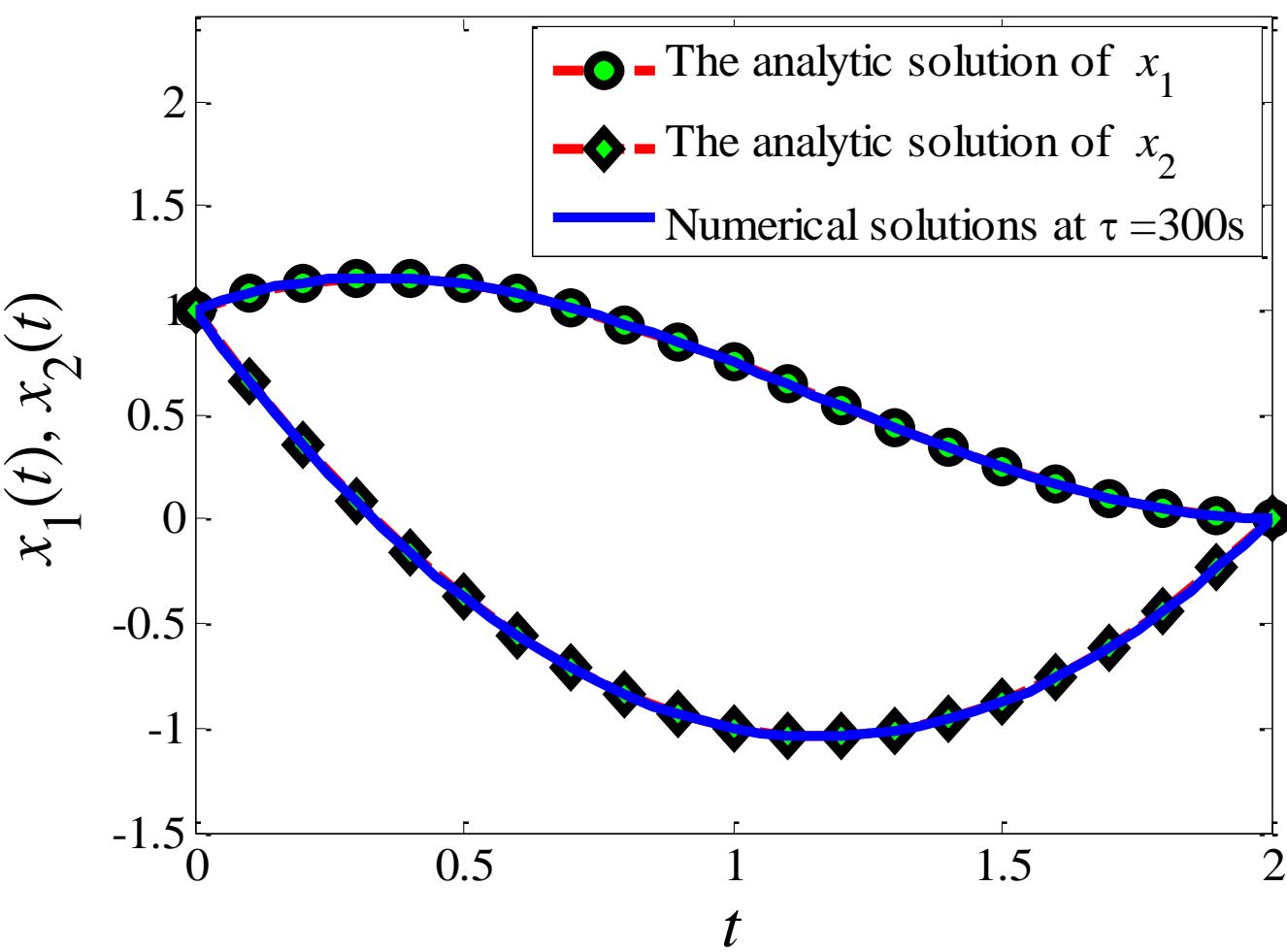

Fig. 3 The numerical solutions of $x_1$ and $x_2$ with the third evolution equation at $\tau = 300\text{s}$ .

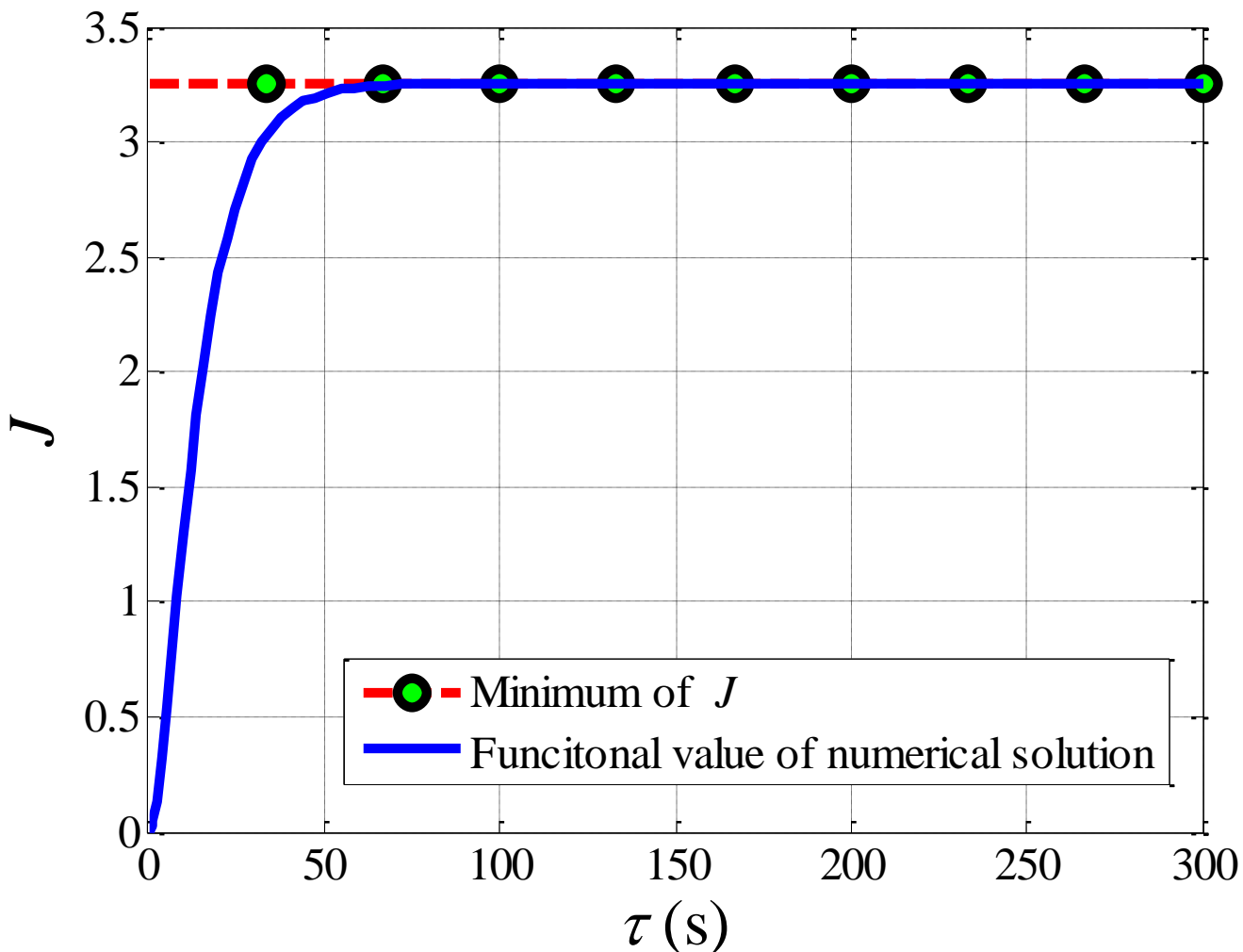

Fig. 4 The approach to the minimum of performance index.

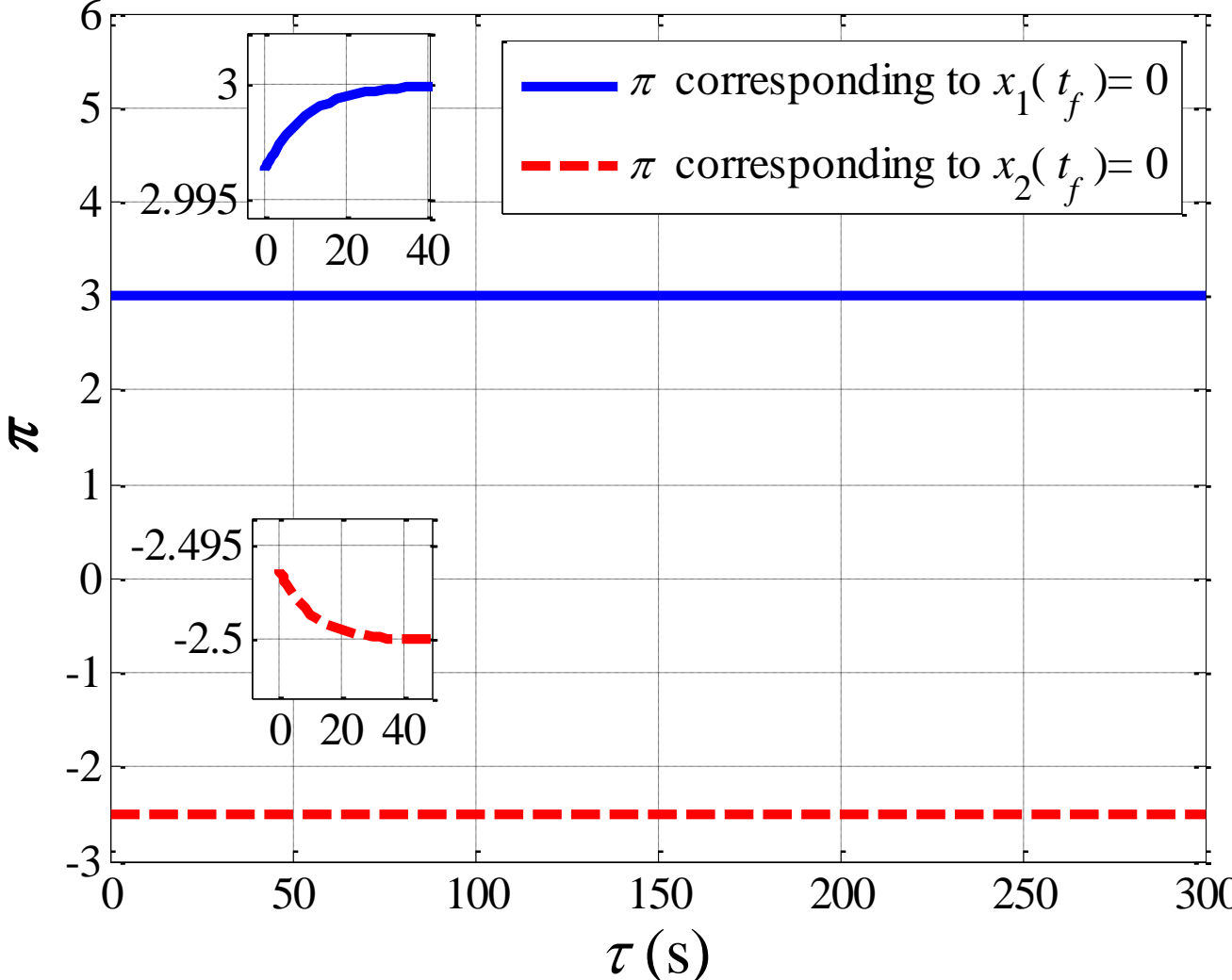

Fig. 5 The evolution profiles of Lagrange multipliers.

Moreover, since both the third evolution equation and the second evolution equation originate from the compact VEM, it makes senses to compare their results. In Table 1, where $e_J = \left| J|_{\tau=300} - \hat{J} \right|$, $e_u = \max\limits_{t\in[t_0,t_f]} \left( \left| u(t,\tau)|_{\tau=300} - \hat{u} \right| \right)$, $e_{x_1} = \max\limits_{t\in[t_0,t_f]} \left( \left| x_1(t,\tau)|_{\tau=300} - \hat{x}_1 \right| \right)$, and $e_{x_2} = \max\limits_{t\in[t_0,t_f]} \left( \left| x_2(t,\tau)|_{\tau=300} - \hat{x}_2 \right| \right)$, the solutions (at $\tau$ =300s) obtained under the same gain parameters and the similar (not optimized) code are compared, and it is shown that the precision is higher with the third evolution equation, while the integration time is slightly larger.

Table 1 Comparison of solutions at $\tau = 300$s between the second and the third evolution equations

| | Number of discretization points | Computation time | $e_J$ | $e_u$ | $e_{x_1}$ | $e_{x_2}$ |
|---|---|---|---|---|---|---|
| The second evolution equation | 123 | 1.68 (s) | $3.7582\times10^{-3}$ | $1.8762\times10^{-3}$ | $1.2008\times10^{-4}$ | $9.3808\times10^{-4}$ |
| The third evolution equation | 41 | 2.84 (s) | $4.8511\times10^{-6}$ | $2.6156\times10^{-6}$ | $2.2578\times10^{-6}$ | $1.5333\times10^{-6}$ |

Now we consider a nonlinear example with free terminal time $t_f$, the Brachistochrone problem [28], which describes the motion curve of the fastest descending.

**Example 2**: Consider the following dynamic system

$$\dot{\boldsymbol{x}} = \boldsymbol{f}(\boldsymbol{x}, u)$$

where $\boldsymbol{x} = \begin{bmatrix} x \\ y \\ V \end{bmatrix}$, $\boldsymbol{f} = \begin{bmatrix} V\sin(u) \\ -V\cos(u) \\ g\cos(u) \end{bmatrix}$, and $g = 10$ is the gravity constant. Find the solution that minimizes the performance index

$$J = t_f$$

with the boundary conditions

$$\begin{bmatrix} x \\ y \\ V \end{bmatrix}\Bigg|_{t_0=0} = \begin{bmatrix} 0 \\ 0 \\ 0 \end{bmatrix}, \quad \begin{bmatrix} x \\ y \end{bmatrix}\Bigg|_{t_f} = \begin{bmatrix} 2 \\ -2 \end{bmatrix}$$

We solved this example with the first evolution equation in Ref. [11] and the second evolution equation in Ref. [14] before. Here it will be addressed via the third evolution equation. In the specific evolution equations (48) and (49), the gain parameters $K$ and $k_{t_f}$ were set to be 0.1 and 0.05, respectively. The $2\times2$ dimensional matrix $\boldsymbol{K}_{\boldsymbol{g}}$ was $\begin{bmatrix} 0.1 & \\ & 0.1 \end{bmatrix}$. The definite conditions were set to be $u(t,\tau)|_{\tau=0} = 0$ and $t_f(\tau)|_{\tau=0}$=1s. We also discretized the time horizon $[t_0, t_f]$ uniformly, with 101 points. Thus, an IVP with 102 states (including the terminal time) is obtained. We still employed "ode45" in Matlab to carry out the numerical integration, for both the IVP of states with respect to $t$ in Eq. (50) and the transformed IVP with respect to $\tau$. In the integrator setting, the default relative error tolerance and the absolute error tolerance were $1\times10^{-3}$ and $1\times10^{-6}$, respectively. The spline interpolation was again used to get the control within the discretization points. For comparison, we computed the optimal solution with GPOPS-II [29], a Radau PS method based OCP solver.

The control solutions are plotted in Fig. 6, and the asymptotical approach of the numerical results are demonstrated. Fig. 7 gives the states curve in the $xy$ coordinate plane, showing that the numerical results starting from the vertical line approach the optimal solution over time. In Fig. 8, the terminal time profile against the variation time $\tau$ is plotted. The result of $t_f$ declines rapidly at first and it then gradually approach the minimum decline time after small oscillation. It is almost unchanged after $\tau$ = 90s. At $\tau$ = 300s, we compute that $t_f$ = 0.8165s, same to the result from GPOPS-II. Fig. 9 presents the evolution profiles of the Lagrange multipliers $\boldsymbol{\pi}$, showing the asymptotically approach to the optimal value of $\begin{bmatrix} -0.1477 \\ 0.0564 \end{bmatrix}$. In Table 2, the optimization results at $\tau$ =300s using the second and the third evolution equations respectively are also compared. Note that the optimal solutions from GPOPS-II are denoted by a hat " ^ ", $e_J = \left| t_f \big|_{\tau=300} - \hat{t}_f \right|$, $e_x = \max_{t\in[t_0,t_f]} \left( \left| x(t,\tau)\big|_{\tau=300} - \hat{x} \right| \right)$, $e_y = \max_{t\in[t_0,t_f]} \left( \left| y(t,\tau)\big|_{\tau=300} - \hat{y} \right| \right)$, and $e_V = \max_{t\in[t_0,t_f]} \left( \left| V(t,\tau)\big|_{\tau=300} - \hat{V} \right| \right)$. It is again shown that the precision of solutions via the third evolution equation is higher, while in this example the computation time consumed is slightly smaller than the second evolution equation, due to the dense discretization granularity.

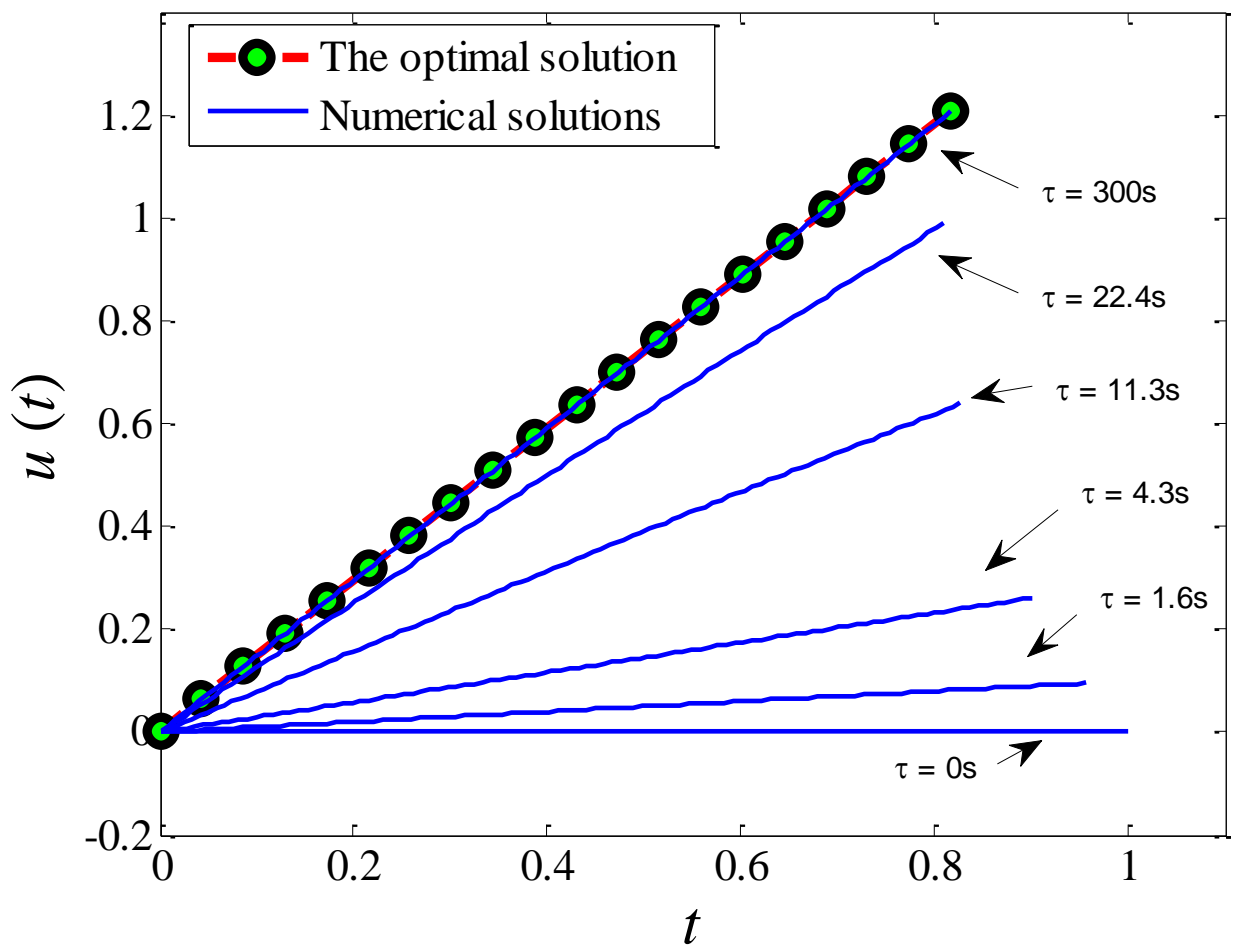

Fig. 6. The evolution of numerical solutions of $u$ to the optimal solution.

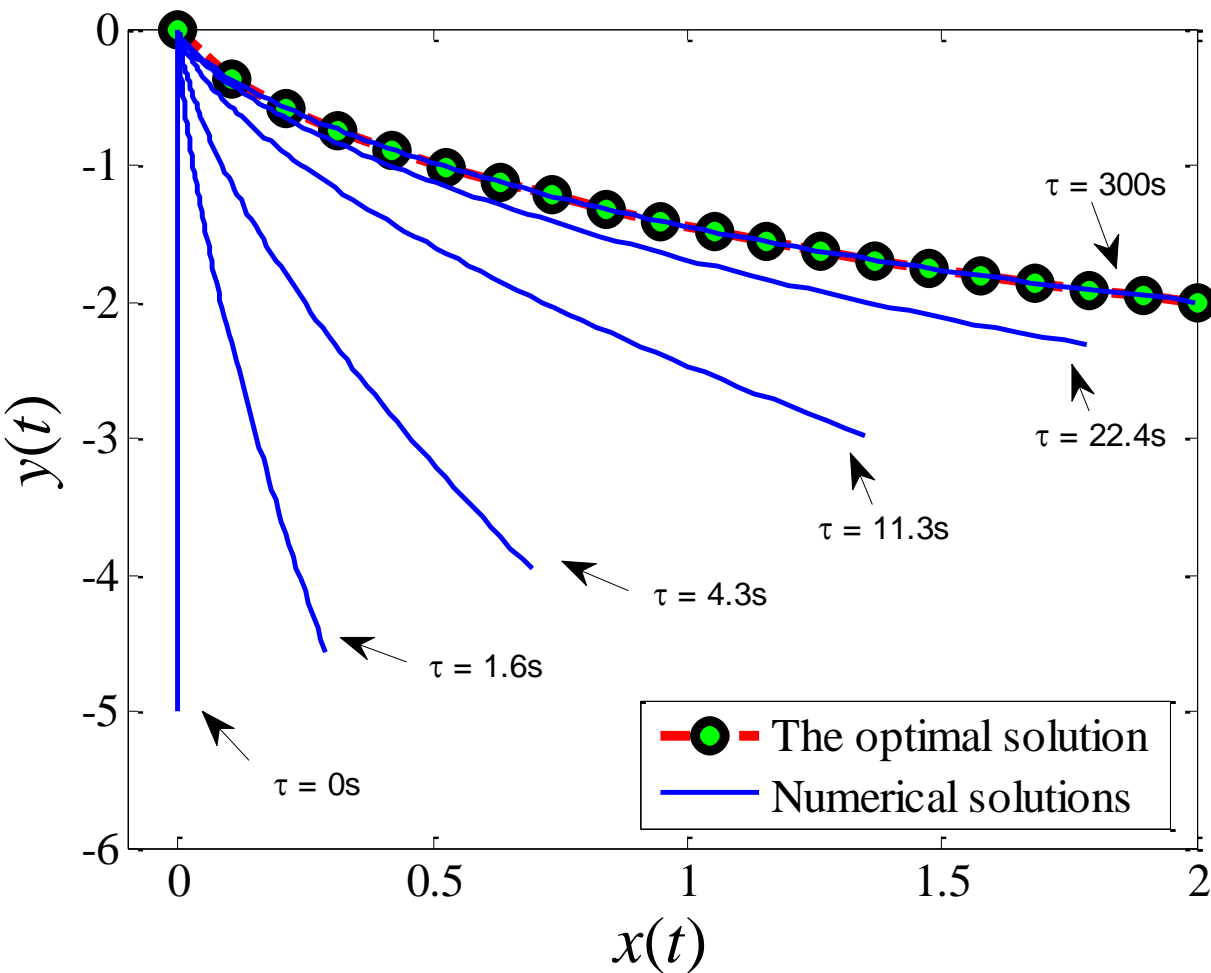

Fig. 7 The evolution of numerical solutions in the $xy$ coordinate plane to the optimal solution.

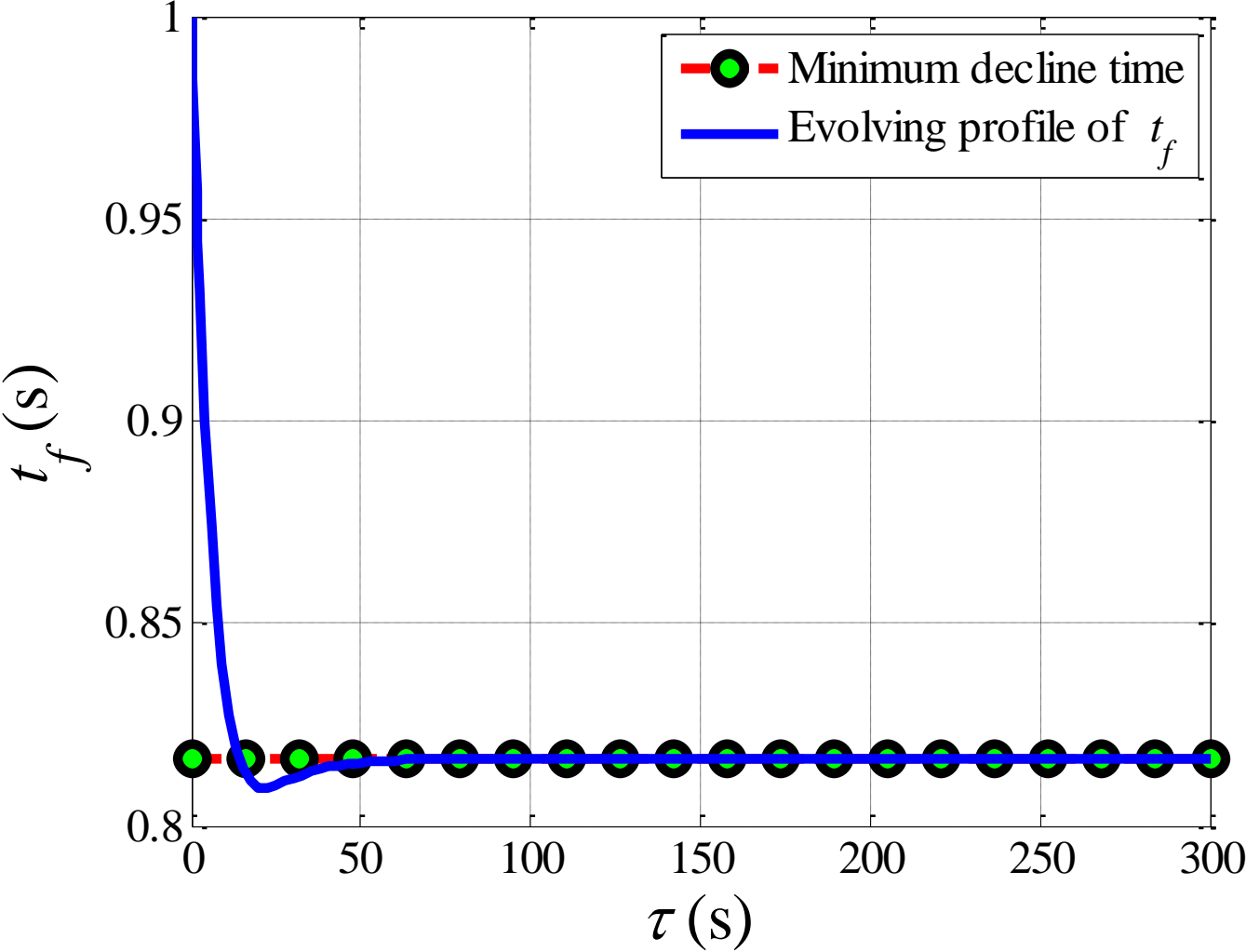

Fig. 8 The evolution profile of $t_f$ to the minimum decline time.

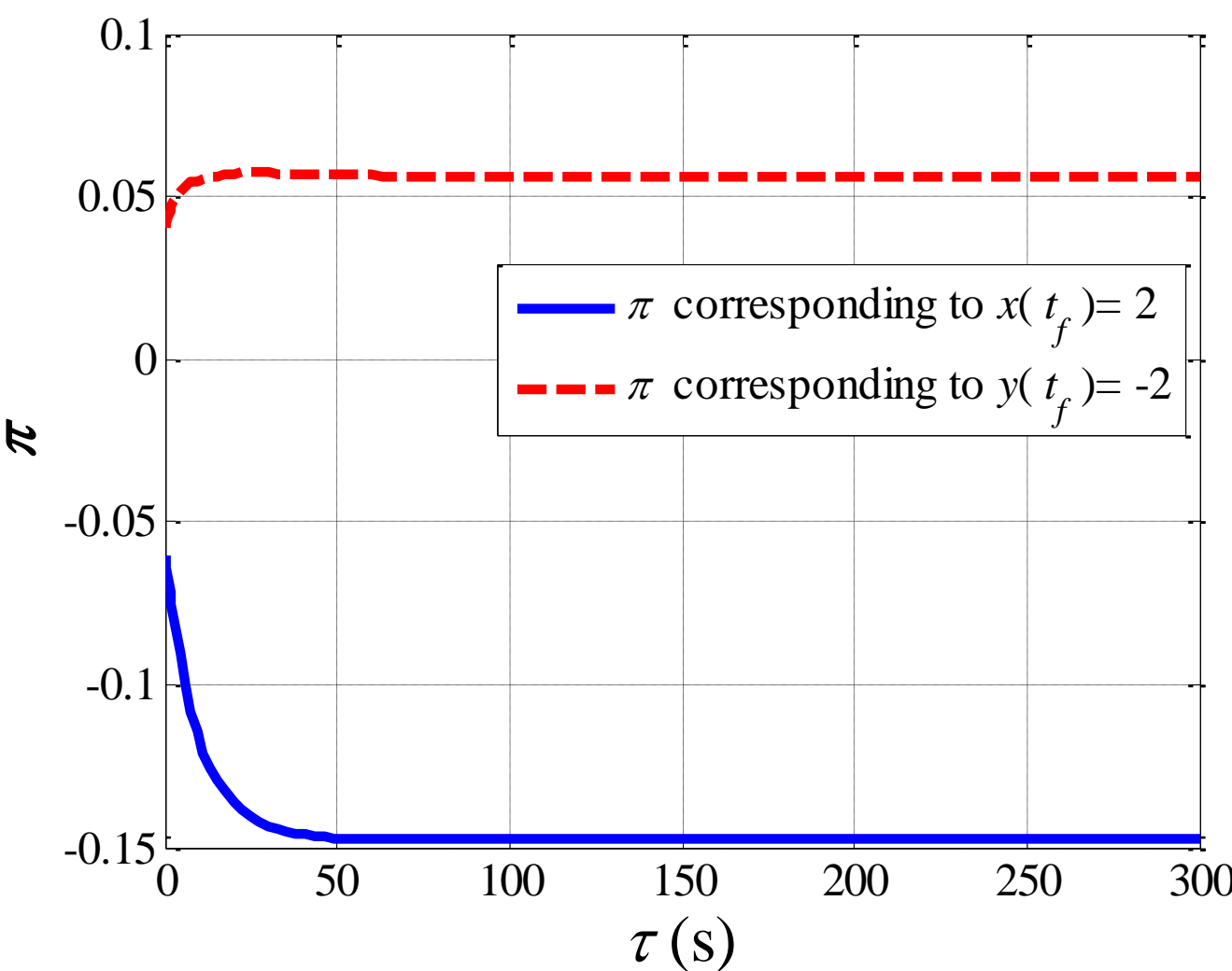

Fig. 9 The evolution profiles of Lagrange multipliers.

Table 2 Comparison of solutions at $\tau = 300\text{s}$ between the second and the third evolution equations

| | Number of discretization points | Computation time | $e_J$ | $e_u$ | $e_x$ | $e_y$ | $e_V$ |
|---|---|---|---|---|---|---|---|
| The second evolution equation | 405 | 13.55 (s) | $3.3841 \times 10^{-5}$ | $1.2297 \times 10^{-3}$ | $2.4491 \times 10^{-4}$ | $6.3434 \times 10^{-3}$ | $3.5051 \times 10^{-3}$ |
| The third evolution equation | 102 | 12.37 (s) | $5.4751 \times 10^{-8}$ | $2.7207 \times 10^{-4}$ | $1.9328 \times 10^{-5}$ | $2.4566 \times 10^{-5}$ | $4.1345 \times 10^{-5}$ |

## V. CONCLUSION

The third evolution equation for the computation of Optimal Control Problems (OCPs) is formulated. It only considers the evolution of the control variables (and the terminal time if needed) along the virtual variation time. Since the states are computable from the state equations (at least numerically), the dimensionality of the Evolution Partial Differential Equation (EPDE) is reduced. When applying the semi-discrete method to solve the EPDE, only the control variables are discretized. Thus the number of states in the transformed Initial-value Problems (IVPs) is greatly reduced, and this might alleviate the computation burden in seeking the numerical solutions. In addition, the computation may be initialized with arbitrary values for the states in the transformed IVP because the equation may address possible infeasibilities. Preliminary tests show that the third evolution equation may give more precise numerical solutions than the second evolution equation, and for the dense discretization case, the computation time used might be less.